\documentstyle{l-aa}    
\input psfig.sty

\def\it{\sl}
 
\begin{document}
 
   \thesaurus{02         
              (01.08.2; 24.01.1) }
   \title{The X-ray lightcurve of SN 1987A}
 
   \author{G. Hasinger\inst{1}, 
           B. Aschenbach\inst{2},
           J. Tr\"umper\inst{2}}
 
   \offprints{G. Hasinger}
 
   \institute{$^1$~Astrophysikalisches Institut Potsdam, 
              An der Sternwarte 16, D--14482 Potsdam, Germany\\
              $^2$~Max Planck Institut f\"ur extraterrestrische Physik,
              Giessenbachstr. 1, D--85740 Garching, Germany}
 
   \date{Received; accepted}
 
   \maketitle
 
   \begin{abstract}
X-ray observations of SN 1987A in the Magellanic Clouds have been performed
throughout the ROSAT mission, using the PSPC and the HRI 
detectors. We present the X-ray light curve based on all 
observations in the years 1991--1995. For the first time a 
significant increase of the X-ray flux from SN 1987A can be 
seen, corresponding to X--ray luminosities (0.5-2 keV) of 
$0.8 - 2.2 \times 10^{34}~erg/s$ about 4 -- 8 years after the explosion. 
SN 1987A is surrounded
by a ringlike nebula, which is thought to form the interface between the 
blue--supergiant wind and the denser red--supergiant wind of the 
progenitor. The X-ray data can constrain the density of the 
matter inside the ring to about $30~amu~cm^{-3}$ and the date at which the 
blast wave will reach the ring to about AD 2003, when a dramatic brightening 
is expected to occur. Nevertheless, other interpretations of the X-ray
emission are possible.
  
\end{abstract}
 
\section{Introduction}

The explosion of SN 1987A (Shelton 1987) was one of the   
exciting historical astronomical events. The neutrino flux
detected at the time of the explosion (Koshiba et al. 1987),
and the observations of X- and soft $\gamma$-rays (Sunyaev et al. 1987,
Dotani et al. 1987)
constitute some of the ``first observations'' of  such phenomena from
supernovae confirming the general correctness of our theoretical
understanding. In addition SN 1987A is classified as type II, 
therefore a neutron star is expected at the center of this remnant.

Hard X-rays have been detected from SN 1987A with the Kvant module on the 
Russian MIR station and with the Japanese Ginga satellite (Sunyaev et al.
1987; Dotani et al. 1987). The high-energy X-rays originate from the
decay of $~^{56}Co$, where the $\gamma$-ray lines undergo multiple Compton
scatterings and are degraded into X-rays which, however, are absorbed 
through the tenuous cocoon of the stellar ejecta which is thick enough
to block out soft X-rays from the compact object for quite a long time.

The ``soft'' X--ray component originally observed with Ginga
below $\sim 10~keV$ could not be confirmed by MIR-TTM and MPE
rocket observations (Aschenbach et al. 1987), nor by the  
ROSAT first-light observations (Tr\"umper et al., 1991). Since
1991, however, the supernova is detected as a faint soft X-ray 
source (Beuermann et al., 1994; Gorenstein et al., 1994). 
In this {\it Letter} we present data from continuous monitoring 
of SN 1987A with ROSAT.

\begin{table}
\caption[ ]{Observation Summary}
\begin{tabular}{lllllll}
\hline
Date & Day$^1$ & Instr. & Time & Count Rate$^2$ \\
     &         &        & [s]  &  [$10^{-3}~c/s$] \\  
\hline
16.06.90-28.06.90 & 1215 & PSPC & ~6400 & $< 2.7^3$       \\ 
11.02.91-13.02.91 & 1448 & HRI  & 23465 & $0.8 \pm 1.0$ \\
21.04.91-07.10.91 & 1645 & PSPC & 24908 & $1.8 \pm 0.5$ \\
03.02.92-14.05.92 & 1872 & PSPC & 27288 & $2.2 \pm 0.5$ \\
04.12.92-05.07.93 & 2258 & PSPC & 23404 & $3.2 \pm 0.6$ \\
28.09.93-30.09.93 & 2408 & PSPC & ~9426 & $3.8 \pm 0.8$ \\
23.12.93-04.10.94 & 2715 & HRI  & 21182 & $3.9 \pm 1.3$ \\
01.01.95-11.10.95 & 3013 & HRI  & 62420 & $4.8 \pm 0.7$ \\
\hline
\end{tabular}

$^1$ mean day after explosion

$^2$ PSPC channel 52--201, HRI channel 2--8 multiplied with 2.65

$^3$ from Tr\"umper et al., 1991

\end{table}

\begin{figure}
\psfig{figure=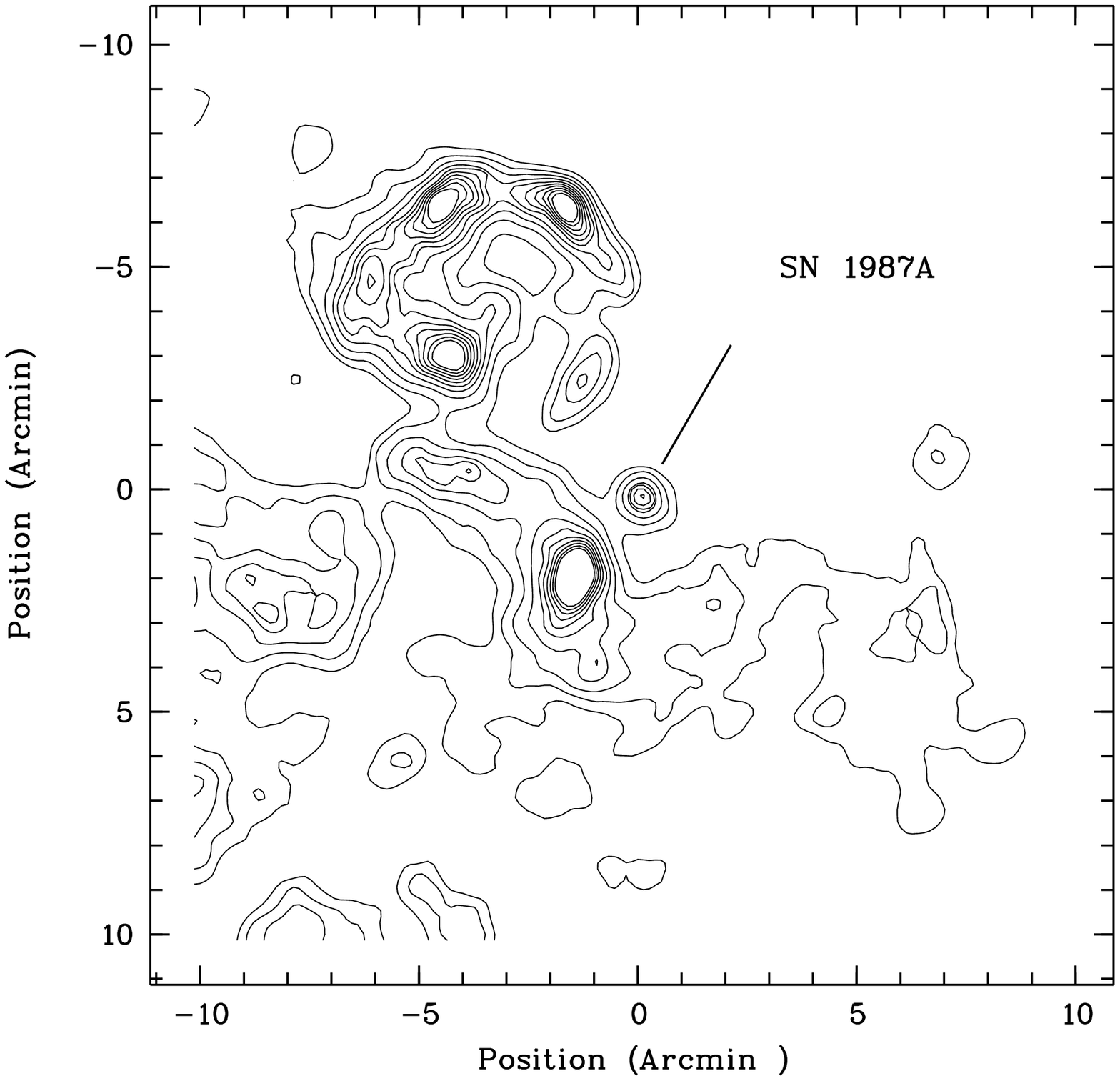,width=8.0cm,bbllx=15mm,bblly=17mm,bburx=205mm,bbury=190mm,clip=}
\psfig{figure=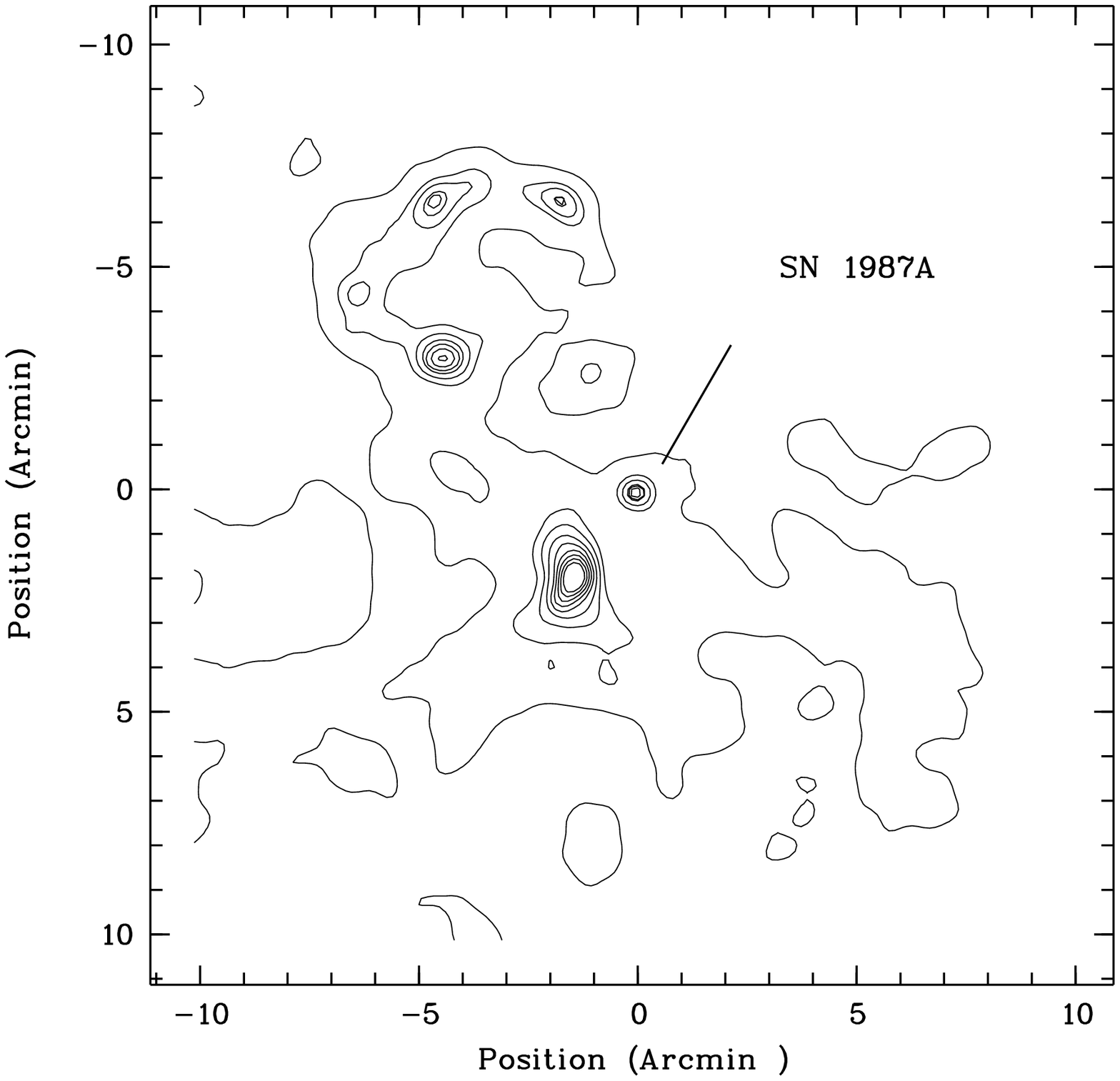,width=8.0cm,bbllx=15mm,bblly=05mm,bburx=205mm,bbury=190mm,clip=}
\caption[ ]{X-ray image of the region of SN 1987A with the
PSPC (top) in the pulse height range 52-201 and the HRI (bottom) in the pulse 
height range 2-8.}
\end{figure}

\section{Observations and Results}

SN 1987A was the target of the ROSAT first light observation, and
was scanned during the ROSAT All-Sky Survey. In both cases only upper limits
could be derived (Tr\"umper et al., 1991). Since then SN 1987A has been 
monitored regularly with both the PSPC and the HRI detectors aboard ROSAT. 
Table 1 gives an overview of the observations. 

For the purpose of this paper all ROSAT data of SN 1987A obtained since 
the beginning of 1991 have been analysed  
using the MPE interactive analysis system EXSAS (Zimmermann et al., 1994).
For the PSPC data all good time intervals selected by the ROSAT 
standard 
analysis (SASS; Voges et al. 1992) have been analysed, yielding  
a net observing time of 85026 seconds. HRI events
have been selected using a custom made procedure after having joined the 
SASS accepted 
and rejected event files. This resulted in a net observing time of 107066
seconds, a gain of $\sim$ 5\% compared to the SASS products. 
Figure 1 shows the X-ray images centered on SN 1987A. The PSPC image 
has been accumulated in the energy range $\sim$ 0.5--2 keV (pulse height 
channels 52-201). The HRI image was restricted to the pulse 
height interval 2--8, accepting a minimal loss ($\sim 6\%$) of  
X-rays but at the same time significantly reducing the intrinsic and 
particle-induced HRI background (see David et al., 1995).
The supernova shows up as a weak point source embedded in a rather 
complex region of diffuse X-ray emission. The average HRI position 
$R.A.(2000)= 5^h35^m28.5"$, $DEC(2000)= -69^o16'11.5"$ is within 2 arcsec
from the catalogued position; no evidence is seen of a possible second,
nearby X-ray source suggested by Gorenstein et al. (1994). 

\begin{table}
\caption[ ]{Spectral Analysis}
\begin{tabular}{lllllll}
\hline
Model$^1$ & $\chi^2_{red}$ & dof & $\Gamma$/kT & Flux$^2$ / EM$^3$ \\
      &          &     & ~[keV]      &  \\  
\hline
power law     & 0.65 & 5 & $3.3 \pm 0.4 $ & $5.38 \pm 0.81$ \\
thermal brems & 0.96 & 5 & $0.44 \pm 0.11$ & $5.36 \pm 0.84$ \\
blackbody     & 1.74 & 5 & $0.17 \pm 0.02$ & $7.88 \pm 2.0 $ \\

RS (z=1)      & 3.61 & 5 & $1.00 \pm 0.16$ & $(5.0 \pm 1.5) \cdot 10^{56}$ \\  
RS (z=0.3)    & 3.00 & 5 & $0.99 \pm 0.17$ & $(1.4 \pm 0.4) \cdot 10^{57}$ \\  
RS (z=3)      & 3.87 & 5 & $1.00 \pm 0.17$ & $(1.8 \pm 0.6) \cdot 10^{56}$ \\  

2T-RS (z=1)   & 1.22 & 3 & 0.14            & $1.8 \cdot 10^{57}$ \\
              &      &   & 1.28            & $6.5 \cdot 10^{56}$ \\
\hline
\end{tabular}

$^1$ $N_H$ fixed to $1.1 \cdot 10^{21}~cm^{-2}$

$^2$ Flux in $10^{-14}~erg~cm^{-2}~s^{-1}$ (PL, TB and BB)

$^3$ Emission measure in $cm^{-3}$ (RS models)

\end{table}

\begin{figure}[htp]
\psfig{figure=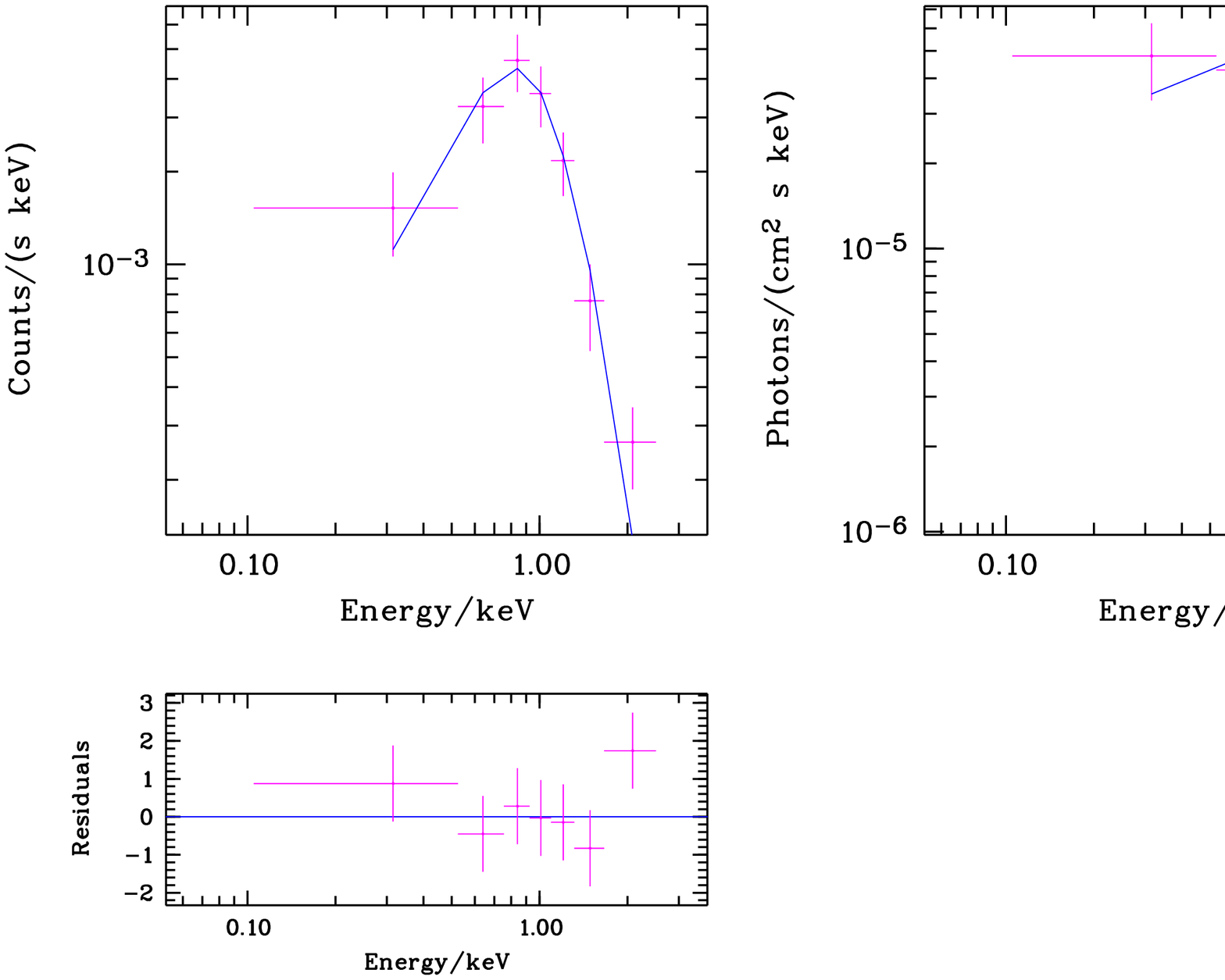,width=8.cm,bbllx=0mm,bblly=58mm,bburx=130mm,bbury=170mm,clip=}
\caption[ ]{PSPC pulse height spectrum of SN 1987A. The solid line shows the 
model for a 0.44 keV thermal bremsstrahlung spectrum and a neutral hydrogen 
column density fixed to $1.1 \cdot 10^{21}~cm^{-2}$.}
\end{figure}

For the spectral analysis care has been taken of the complex diffuse 
emission in the vicinity of the supernova. PSPC source photons were accumulated 
in a circle around SN 1987A with radius 36", while background photons were 
taken from an annulus with inner radius 36" and outer radius 82". 
The pulse height distribution in the range 11-250 channels has been binned 
to a signal to noise ratio of 4, yielding 318
net counts. Various spectral models have been fit to the data, assuming
a fixed line-of-sight absorption column density of $N_H = 10^{21}~cm^{-2}$
(Gorenstein et al., 1994). 
The PSPC response matrix is known to change as a function of time
(Prieto et al., 1996), a fact which is taken care of by EXSAS using
different instrument matrices for the beginning and the end of the
PSPC lifetime. Since the PSPC data covers a long period of time 
(1991 Apr - 1993 Sep) the fits were performed with both response matrices
in order to test for systematic errors, however no significant
differences were found. Table 2 gives an overview of the spectral fits. 
Continuum spectra like a simple power law with photon index 
$\Gamma = 3.3$ or thermal bremsstrahlung with $kT = 0.44~keV$
fit the data well. Figure 2 shows the 
PSPC pulse height spectrum with the corresponding thermal bremsstrahlung model.
The average 0.5-2 keV luminosity derived for the power law or bremsstrahlung 
fit is $(1.6\pm0.2)\cdot10^{34}~erg~s^{-1}$, assuming a distance of 51 kpc
to the LMC. A single-temperature Raymond-Smith 
thermal plasma model (RS), which is dominated by few emission lines, is  
rejected at a probability greater than 99\%, even if the  
abundance is varied. This may indicate a more complicated ionization or
temperature distribution. 
A two-temperature Raymond-Smith model is given for completeness, however, 
the errors are too large to draw firm conclusions.

\begin{figure}
\psfig{figure=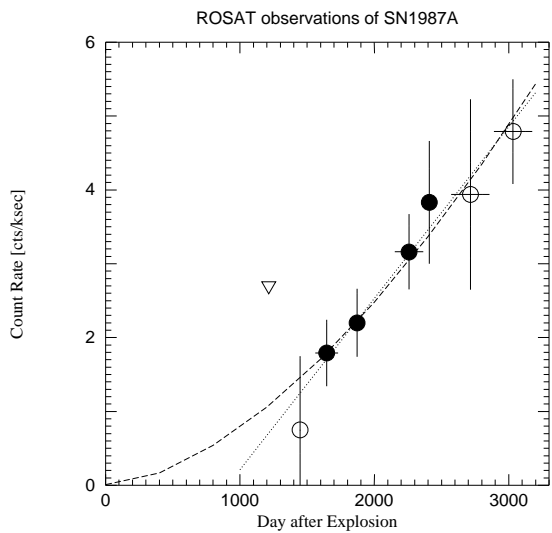,width=8.0cm,bbllx=20mm,bblly=70mm,bburx=210mm,bbury=260mm,clip=}
\caption[ ]{Time evolution of the X-ray emission of SN 1987A
The data points marked with filled circles correspond to 
PSPC observations (channel 52-201), open circles are from HRI observations
(channel 3-8) multiplied by a factor of 2.65. The dashed line indicates 
a quadratic fit to the data. A linear increase starting on day $\sim 900$
can fit the data as well (dotted line).}
\end{figure}

In order to obtain a light curve of SN 1987A, separate images 
were produced
for all time intervals in table 1. Source counts were extracted from 
a ring around the supernova with radius 24" for the HRI and 40" for the
PSPC, respectively, containing roughly 90\% of the HRI (David et al., 
1995) and about 95\% of the PSPC photons (Hasinger et al., 1992), respectively. 
Background counts were extracted from an annulus around the source ring, having
an outer radius of 82". Net source count rates were corrected for deadtime 
and the point-spread-function losses. In order to compare the HRI and PSPC
count rates we computed the ratio of the count-to-flux conversion factors
for both instruments, taking into account the selected 
pulse height channel range (channel 52-201 for the PSPC and channel 2-8 for 
the HRI). Assuming the first three spectral models in table 2, the predicted 
count rate ratio factor varies between 2.46 and 2.89. We therefore assumed
a PSPC/HRI conversion factor of 2.65. The final equivalent PSPC count rates
are given in table 1. These count rates can be converted to 0.5-2 keV 
fluxes assuming a factor of $6.95 \cdot 10^{10}$ PSPC counts per 
$erg~cm^{-2}~s^{-1}$. Figure 3 shows the X-ray light curve of SN 1987A, 
clearly indicating a continuous flux increase during the interval 1500--3000
days after the explosion. We fitted the count rates with a power law model as
a function of time and derive an exponent of $1.67\pm0.35$ ($1\sigma$).
A linear increse starting on day $\sim 900$ can fit the data as well.

\section{Discussion}

Using the FOC aboard the Hubble Space Telescope
a faint circumstellar nebula has been observed around SN 1987A which
has the shape of a thin ring with a radius of 0.2 pc and a thickness
of only 0.02 pc (Jakobsen et al. 1991). The authors estimate a density of 
$n_e \sim 2 \cdot 10^4 cm^{-3}$ and a total mass of 
$ M \sim 0.2 M_\odot$ for the ring, which is thought to originate
from a shock front in the progenitor wind at the interface between a slow, 
tenuous red supergiant wind (RSW) and a fast asymmetrical low--density 
blue supergiant wind (BSW). In the meantime high quality images of the
Hubble Space Telescope show two fainter rings above and below the
original ring nebula which probably mark the hour-glass shaped
transition region between the BSG and the RSG winds, where a  substantial 
density enhancement is expected to occur. The ring nebula is also 
observed to expand very slowly, presumably somewhat faster than the RSG 
wind speed (Plait et al., 1995).

Several investigators have discussed the X-ray emission of
supernovae, in particular that of SN 1987, in terms of thermal radiation of 
a reverse shock wave originating from the interaction between the supernova 
ejecta and the previously existing circumstellar material (e.g. Chevalier 1982).
The supernova blast wave is expected to interact with the wind, first with
the BSG and later with the RSG wind, causing 
a shock wave to heat up the circumstellar material to X-ray temperatures.

The X-ray luminosity of the shock wave and its temporal behaviour depend
on the density and gradient of the ambient medium. In the case of
a freely expanding BSG wind with a density  
$\rho \sim r^{-2}$ a decline of the X-ray luminosity as a function of 
time according to $L_X \sim t^{-0.87}$ is expected (Beuermann et al., 1994;
Gorenstein et al., 1994). Beuermann et al. have already remarked that
the measured X-ray luminosity is too high for the simple
free wind model. Our current X-ray light curve is clearly
inconsistent with the free wind model.  

However, there are reasons to believe that a simple $r^{-2}$
density dependence is not the correct description for
the BSG wind of SN 1987A. As the existence of the ring nebula
and the terminating discontinuity prove, the BSG wind must catch
up and interact with the RSG wind and is therefore no longer
freely expanding. 
Several authors recently assumed the BSG wind matter to be homogeneously
distributed within the radius of the ring nebula
(Suzuki et al., 1993; Masai and Nomoto; 1994; Luo et al., 1994). 
In this case a
steadily increasing X-ray luminosity is predicted in the BSG wind
interaction phase, with a time dependence proportional
to $t^{2.04}$, which is consistent with our findings (see fig. 3). 
Chevalier \& Dwarkadas (1995) propose that the supernova shock 
is now interacting with an $H_{II}$ region created by the BSG star
in the swept-up RSG wind. They predict a slow rise of the X-ray
flux, which, however should start only around day $\sim 1000$.
Our supernova X-ray light curve is consistent with this model, too.

Masai \& Nomoto (1994)  
present detailed hydrodynamical calculations and predict
a dramatic increase in X-ray luminosity around the 
time when the stellar ejecta hit the dense ring nebula. 
Because the SN blast wave is slowed down by the shock interaction, 
the impact time depends on the density of the ambient medium according
to $ t_{ring} \sim \rho^{1/6}$. Different authors assume substantially
different values for this density, ranging from 
$\rho \approx 2~amu~cm^{-3}$ (Luo et al., 1994; Masai \& Nomoto, 1994) to
$\rho \approx 100~amu~cm^{-3}$ in the equatorial plane
(Chevalier \& Dwarkadas, 1995).  
Correspondingly, the range of predicted impact dates is still considerable
(roughly 1996 -- 2008). The new ROSAT determination of the X-ray
lightcurve of SN 1987A now allows to put further constraints on the
density in the circumstellar material. The total X-ray flux measured on
day 3000 (see fig. 3 and tab. 1) corresponds to a 0.5-2 keV luminosity
of $L_{(0.5-2)} = 2.0 \cdot 10^{34}~erg~s^{-1}$. Depending on the assumed 
spectral
model, the energy band correction factor is quite uncertain. Assuming
a correction factor of 3 for a thermal bremsstrahlung spectrum we estimate a 
total X-ray luminosity 
of $L_X = 6 \cdot 10^{34}~erg~s^{-1}$. If we interpret this 
as coming from the SN blast wave in the context of the Masai \& 
Nomoto (1994) model, we can estimate the density of the BSG wind to
$\rho_{BSW} \approx 30~amu~cm^{-3}$ from their equation (7). Correspondingly, 
the impact time of the SN blast wave at the location of the ring nebula 
would be $t_{ring} \approx 15.7~yr$ after the explosion, i.e. in the
year AD 2003. We have to caution, however, that we find significantly 
lower temperatures than predicted by Masai \& Nomoto. A factor of $\sim 3$ 
higher density and, correspondingly,
a later impact time has been proposed by Chevalier and Dwarkadas (1995).
In view of the rather complicated situation these predictions are still
quite uncertain.

There could be other contributions to the observed X-ray emission of 
SN 1987A, in particular radiation from the surface or the magnetosphere 
of a putative neutron star. Actually, the 
currently observed X-ray luminosity and temperature are  
comparable to the values expected from the thermal surface radiation of a young
neutron star. However, the circumstellar cocoon of matter is expected
to block out all but the highest energy X-rays of the neutron star for a long 
time. Under relatively optimistic assumptions, Kumagai et al. (1993) predict
a possible pulsar to be seen after $\sim 8~yr$ only at energies above 6 keV.    
Nevertheless, relatively little is known about the clumpiness of the 
absorber and thus the observed increase of X-ray flux could still be 
associated to radiation from a compact object. The long-term X-ray lightcurve
of SN 1987A can likely discriminate between the different possibilities.
Continuing X--ray observations of SN 1987A are therefore highly desirable,
in particular since at any rate a major brightening is expected in the 
coming years.

   \acknowledgements
We would like to thank K. Beuermann, R. McCray, K. Masai, K. Nomoto and 
H. \"Ogelman for important communications. We thank an anonymous referee
for helpful comments. The ROSAT project is supported by 
the Bundesministerium f\"ur Bildung, Wissenschaft und Forschung (BMBF) 
and by the Deutsche Agentur f\"ur Luft und Raumfahrt (DARA).

\end{document}